\documentclass[%
 reprint,
 amsmath,amssymb,
 aps,
 pre,
floatfix, aip,numerical,
]{revtex4-1}

\usepackage{graphicx}
\usepackage{dcolumn}
\usepackage{bm}
\usepackage{verbatim}
\usepackage{amsmath}
\usepackage{mathtools}
\usepackage{color}

\begin{document}


\title{Self-assembly of three-dimensional ensembles of magnetic particles with laterally shifted dipoles}

\author{Arzu B. Yener}
\author{Sabine H. L. Klapp}%
 \email{klapp@physik.tu-berlin.de}
\affiliation{%
 Institute of Theoretical Physics, Technical University Berlin, 10625 Berlin, Germany
}%




\date{\today}
%
\begin{abstract}
We consider a model of colloidal spherical particles carrying a permanent dipole moment which is laterally shifted out of the particles' geometrical centres, i.e. the dipole 
vector is oriented perpendicular to the radius vector of the particles. Varying the shift $\delta$ from the centre, we analyze  ground state structures for 
two, three and four hard spheres, using a simulated annealing procedure. We also compare to earlier ground state results. We then consider a bulk system at finite temperatures and different densities. Using Molecular Dynamics simulations, we examine the equilibrium self-assembly properties for several shifts. 
Our results show that the shift of 
the dipole moment has a crucial impact on both, the ground state configurations as well as the self-assembled structures at finite temperatures. 
%
%
\end{abstract}
\pacs{Valid PACS appear here}
\maketitle
%
%
\section{\label{intro}Introduction}
Recent advances in particle synthesization and the permanent need for novel materials meeting more and more specialized requirements encourage the search for novel types 
of functionalized particles. Promising candidates in this area are colloids with directional interactions. These interactions are the key for the controlled 
self-assembly of colloidal particles into specific structures. Recent research on this topic resulted in complex colloidal particles 
characterised by complex shape~\cite{liddell,bibette,wang_nature}, anisotropic internal symmetry~\cite{glotzer,wang_langmuir} or surface charges~\cite{kegel,chen}. Understanding
the interaction-induced behaviour of such particles is crucial for optimizing their application e.g. in material science biomedicine or sensors.\\
Yet, not only the application of functionalized particles is of interest but also their capability to serve as 
model systems to study fundamental concepts of physics such as self-organization~\cite{granick,pine}, chirality~\cite{bibette, schamel, jampani,ma}, 
synchronisation~\cite{juniper,kotar,granick,jaeger}, 
critical phenomena~\cite{verso,vink}, entropic effects~\cite{mao,anders} and active motion~\cite{ebbens,kogler}, to name a few. 
A paradigm example is the model of dipolar hard and soft spheres which is a well-established model to examine and understand 
the properties of magnetic colloidal particles immersed in a solvent, also called ferrofluids. From numerous studies of the phase behavior of dipolar liquids
 (e.g.~\cite{Hyninnen,RovigattiRusso}), and especially 
of their structural properties (e.g.~\cite{RovigattiSciortino,KantorovichSciortino}), it is known that the dipoles assemble into chains, 
rings and branched structures at sufficiently large dipolar strengths and low densities.\\
\indent Here, we focus on (permanent) ferromagnetic colloids with anisotropic symmetry, i.e. the magnetic moment within the colloidal particle is not located at the geometric
centre of the particle. A first theoretical description for decentrally located dipoles was introduced by 
Holm {\it{et al.}}~\cite{holm1, holm2,Holm_cluster}. In their model, spherical particles carry a dipole moment which is shifted out of the particle centre and is oriented 
parallel to the raduis vector of the 
particle. The model describes very well the cluster formation of particles carrying a magnetic cap~\cite{Baraban}. Yet, it is insufficient to mimic the self-assembly of
so-called Patchy colloids~\cite{pine}, that is, silicon balls carrying magnetic cubes beneath their surfaces. Furthermore, the model 
does not reproduce the zig-zag chained structures formed by magnetic Janus particles in an external field, i.e. particles where one hemisphere of silica spheres are 
covered with a magnetic coat~\cite{granick,kretchmar}. The concept of shifting the dipole was later extended by fixing the amount of shift and varying the orientation of the dipole 
moment vector within the particle~\cite{Abrikosov} which was proven to be more convenient for patchy collids.\\
In the present contribution we consider a model in which the dipole moment is laterally shifted such that the radius vector and the dipole moment vector are oriented 
perpendicular. The same model was also proposed in Ref.~\cite{Abrikosov}.However, here we fix the orientation of the dipole moment and vary the amount of shift. Thereby, we do not only 
aim at modeling synthesized particles mentioned above. Rather, we are interested in understanding the impact of successively shifting the dipole on the self-assembly of such 
particles. To this end, we perform ground state calculations for a small number of dipolar hard spheres and conduct Molecular Dynamics (MD) simulations to study the bulk at 
finite temperature in three dimensions. Very recently, Novak {\it{et al.}} have considered the same model~\cite{novak}, however, they restricted their study to systems
where the particles are fixed in a plane with freely rotating dipoles. Thus, they considered a quasi-twodimensional (q2D) system. Besides, the authors examined the system at 
one fixed density. Here, we examine a three dimensional system of such particles at zero temperature and conduct MD simulations of the bulk at several thermodynamic state 
points. Thereby, we aim at a quantitive characterization in which the shift is the stateparameter in the system.\\
The remainder of the paper is structured as follows. In section~\ref{model}, we present the model and the equations of motion, sec.~\ref{comp} refers to the computaitonal 
methods and sections~\ref{prelim} and~\ref{bulk} include the results for the ground state calculations and for the structural analysis of the bulk systems, 
respectively. We close the paper by a summary and outlook.\\
\section{\label{model}Model and Equations of motion}
\begin{figure} 
\begin{minipage}{9cm}
\includegraphics[scale=1.2]{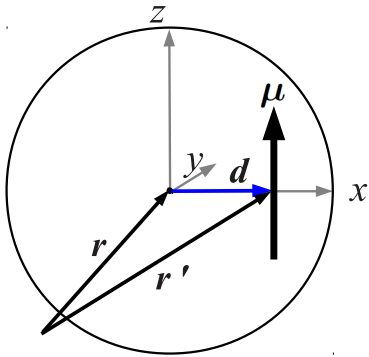}     
\end{minipage}
\caption{\label{skizzen} Sketch of a dipolar sphere with a laterally shifted dipole moment. Also shown are the axes of the body-fixed 
coordinate system.}
\end{figure}

Our model consists of $N$ spherical particles carrying a permanent dipole moment $\boldsymbol{\mu}_i$, ($i=1,..,N$), which is laterally shifted with respect to the 
center of particles. A sketch is given in Fig.~\ref{skizzen}. In the body-fixed reference frame (in the following denoted by the subscript $b$), the location of $\boldsymbol{\mu}_i$ is specified by the shift vector 
$\boldsymbol{d}^b_i = d\,(1,0,0)$,  and its orientation is given by 
the vector $\boldsymbol{\mu}^b_i = \mu\,(0,0,1)$, with $d$ and $\mu$ being constant for all particles. 
Hence, $\boldsymbol{d}_i$ and $\boldsymbol{\mu}_i$ are oriented perpendicular to one another. Thus, our particles differ from those considered 
in Ref.~\cite{holm2} where $\boldsymbol{d}_i$ and $\boldsymbol{\mu}_i$ are arranged parallel and hence $\boldsymbol{\mu}_i$ is shifted radially.\\ 
In the laboratory reference frame, $\boldsymbol{r}_{i}$ is the position vector of the particle centre while the position vector of $\boldsymbol{\mu}_i$ is given by 
$\boldsymbol{r}^{\prime}_{i} = \boldsymbol{r}_{i} + \boldsymbol{d}_i$, where $\boldsymbol{d}_i$ now denotes the shift vector in the laboratory frame. For $d = 0$, 
$\boldsymbol{r}'_{i}$ coincides with $\boldsymbol{r}_{i}$ yielding conventional dipolar systems with centered dipoles. 
The total pair potential between two particles $i$ and $j$ consists of a short-range repulsive potential, $u_{short}(r_{ij})$, and the dipole-dipole potential, 
\begin{center}
\begin{equation} \label{uDD}
 u_{DD}(i,j) = \frac{\boldsymbol{\mu}_i \cdot \boldsymbol{\mu}_j}{{r^\prime}_{ij}^3} - \frac{3(\boldsymbol{\mu}_i \cdot \boldsymbol{r^\prime}_{ij})(\boldsymbol{\mu}_j \cdot \boldsymbol{r^\prime}_{ij})}{{r^\prime}_{ij}^5},
\end{equation}
\end{center}
yielding
\begin{center}
\begin{equation} \label{uij}
 u(i,j) = u_{short}(r_{ij}) + \frac{\boldsymbol{\mu}_i \cdot \boldsymbol{\mu}_j}{{r^\prime}_{ij}^3} - \frac{3(\boldsymbol{\mu}_i \cdot \boldsymbol{r^\prime}_{ij})(\boldsymbol{\mu}_j \cdot \boldsymbol{r^\prime}_{ij})}{{r^\prime}_{ij}^5}.
\end{equation}
 \end{center}
 Here, $r_{ij} = |\boldsymbol{r}_{ij}| = |\boldsymbol{r}_{i} - \boldsymbol{r}_{j}|$ is the center-to-center distance of particles $i$ and $j$, while  
 $r^{\prime}_{ij} = |\boldsymbol{r^{\prime}}_{ij}| = |\boldsymbol{r}_{ij} + \boldsymbol{d}_{ij}|$, with $\boldsymbol{d}_{ij} = \boldsymbol{d}_i - \boldsymbol{d}_j$, 
 determines the distance between the dipoles. 
 We employ two different types of repulsive interactions. First, for the finite temperature MD simulations discussed in Sec.~\ref{bulk}, the repulsive potential is modeled 
 via the shifted soft sphere (SS) potential defined as   
\begin{center}
\begin{equation} \label{usoft}
 u_{SS}(r_{ij})=\epsilon\big(\frac{\, \sigma}{r_{ij}}\big)^n - \big(\epsilon\frac{\, \sigma}{r_c}\big)^n + (r_c-r_{ij})\,\frac{d\,u_{SS}}{dr_{ij}}(r_c).
 \end{equation}
\end{center}
The parameters for potential depth and steepness, $\epsilon$ and $n$, respectively, are specified in Sec.~\ref{bulk}. At the cut-off distance $r_c$, the shifted potential 
given in Eq.~(\ref{usoft}) and its first derivative continuously go to zero such that corrections due to the cut-off are not required. Finally, $\, \sigma$ represents the diameter 
of the particles.\\
Second, for the ground-state calculations presented in Sec.~\ref{prelim}, we set $u_{short}(r_{ij})$ equal to the hard sphere (HS) potential defined as 
\begin{center}
\begin{eqnarray} \label{uhard}
 u_{HS}(r_{ij}) = \left\{\begin{array}{ll}\infty,  & r_{ij} \le \sigma \\
         0, & r_{ij} > \sigma \end{array} \right.  \, .
 \end{eqnarray}
\end{center}
%
We now derive the equations of motion of the particles in the absence of a solvent. Each particle $i$ experiences the total force 
$\boldsymbol{F}_i = {\boldsymbol{F}_i}^{short} + {\boldsymbol{F}_i}^{DD}$ at its centre of mass. The force 
\begin{center}
\begin{eqnarray} \label{eom_trans}
{\boldsymbol{F}_i}^{short} = -\sum_{j\ne i} \boldsymbol{\nabla}_{\boldsymbol{r}_{ij}} u^{short}(r_{ij})
\end{eqnarray}
\end{center}
is due to steric interactions with all other particles $j$ and 
\begin{center}
\begin{eqnarray} \label{eom_trans}
{\boldsymbol{F}_i}^{DD} = - \sum_{j\ne i} \boldsymbol{\nabla}_{\boldsymbol{r}'_{ij}} u^{DD}(\boldsymbol{\mu}_i, \boldsymbol{\mu}_j, \boldsymbol{r}'_{ij})
\end{eqnarray}
\end{center}
is the dipolar force due to the dipole-dipole potential $u^{DD}(\boldsymbol{\mu}_i, \boldsymbol{\mu}_j, \boldsymbol{r}'_{ij})$ given in Eq.~(\ref{uDD}). 
Note that although the force ${\boldsymbol{F}_i}^{DD}$ acts at $\boldsymbol{r'}_i$ , the same force also acts on the center $\boldsymbol{r}_i$ due to the rigidity of 
the particle. Moreover, the finite shift $\boldsymbol{d}_i$ generates a torque ${\boldsymbol{T}_i}^{d} = \boldsymbol{d}_i \times {\boldsymbol{F}_i}^{DD}$ acting at 
$\boldsymbol{r}_i$, which supplements the torque ${\boldsymbol{T}_i}^{\mu} = \boldsymbol{\mu}_i \times {\boldsymbol{G}_i}^{DD}$ stemming from angle dependent dipolar 
forces~\cite{klappbuch}. Here,
\begin{center}
\begin{eqnarray} \label{eom_trans}
{\boldsymbol{G}_i}^{DD} = - \sum_{j\ne i} \boldsymbol{\nabla}_{\boldsymbol{\mu}_i} u^{DD}(\boldsymbol{\mu}_i, \boldsymbol{\mu}_j, \boldsymbol{r}'_{ij})\,.
\end{eqnarray}
\end{center}
Thus, the total torque on the particle centre is given by $\boldsymbol{T}_i = {\boldsymbol{T}_i}^{\mu} + {\boldsymbol{T}_i}^{d}$. For $d=0$, i.e. 
$\boldsymbol{r'}_i = \boldsymbol{r}_i $,
the additional torque ${\boldsymbol{T}_i}^{d}$
vanishes and the forces and torques reduce to the expressions familiar for centered dipoles (e.g.~\cite{tildesley}).
We also note that our treatment of the forces and torques in a system of shifted dipoles is equivalent to the virtual sites method introduced by Weber {\it{et al.}}~\cite{holm2}.
Having derived the forces and torques, the (Newtonian) equations of motion
are given by
\begin{center}
\begin{eqnarray} \label{eom_trans}
m\ddot{\boldsymbol{r}}_i = \boldsymbol{F}_i 
\end{eqnarray}
\end{center}
for translation (with $m$ beeing the mass of the particles), and
\begin{center}
\begin{eqnarray} \label{eom_rot}
&{\boldsymbol{T}_i}^b &= I \dot{{\boldsymbol{\omega}}}_i^b\\
&{\dot{\boldsymbol{Q}_i}} &= \frac{1}{2} \boldsymbol{W}{\boldsymbol{\Omega}_i}^b \label{eom_rot2} 
\end{eqnarray}
\end{center}
for rotation~\cite{tildesley}. In Eq.~(\ref{eom_rot}), ${\boldsymbol{\omega}}_i$ is the angular velocity and $I$ is the moment of inertia. Further, quantities in the body 
fixed frame can be transformed to the laboratory frame via a rotation matrix given in~\cite{tildesley}. 
In Eq.~(\ref{eom_rot}), the quantity $\dot{{\boldsymbol{Q}_i}}$ is the time derivative of the quaternion
${\boldsymbol{Q}_i} = ({q_i}^0, {q_i}^1, {q_i}^2, {q_i}^3)$ which we employ to describe the orientation of the particle (specified in~\cite{tildesley}). The matrix 
$\boldsymbol{W}$ is defined as (see~\cite{tildesley})
\begin{equation}
 W = \begin{pmatrix}
q_i^0 & -q_i^1 & -q_i^2 & -q_i^3 \\
q_i^1 &  q_i^0 & -q_i^3 &  q_i^2 \\
q_i^2 &  q_i^3 &  q_i^0 & -q_i^1 \\
q_i^3 & -q_i^2 &  q_i^1 &  q_i^0
\end{pmatrix}
\end{equation}
while the quaternion ${\boldsymbol{\Omega}_i}^b = (0, {\omega_{ix}}^b, {\omega_{iy}}^b, {\omega_{iz}}^b )$
corresponds to the $x$, $y$ and $z$ components of the angular velocity. It can be shown that Eq.~(\ref{eom_rot2}) is equivalent to the expression 
$\dot{\boldsymbol{s}}_i = \boldsymbol{\omega}_i \times \boldsymbol{s}_i$ known for the rotation of linear molecules (e.g.~\cite{tildesley}), where $\boldsymbol{s}_i$ is 
the unit vector of the particle orientation and $\dot{\boldsymbol{s}}_i$ its time derivative.
\section{\label{comp}Computer simulations}
\subsection{\label{md}Molecular Dynamics simulation}
In our MD simulations, we constrain the temperature to a constant value $T$ by using a Gaussian isokinetic thermostat~\cite{tildesley}. Hence, the 
linear and angular momenta of the particles are rescaled by the factors $\mathcal{X}_{trans} = \sqrt{T/\mathcal{T}_{trans}}$ and $\mathcal{X}_{rot}= \sqrt{T/\mathcal{T}_{rot}}$, 
respectively, where $\mathcal{T}_{trans} = 1/(3Nk_BT)\sum_{i=1}^N m {\dot{r_i}}^2$ (with ${\dot{r_i}} = |d\boldsymbol{r}_i/dt|$) and 
$\mathcal{T}_{rot} = 1/(3Nk_BT)\sum_{i=1}^N I{\omega_i}^2$ (with ${\omega_i} = |\boldsymbol{\omega}_i|$) are the translational and rotational 
kinetic temperatures of the system. Further, $k_B$ is the Boltzmann constant.
We solve the corresponding isokinetic equations for translational and rotational motion with a Leapfrog algorithm, following the schemes suggested in Refs.~\cite{tildesley} and~\cite{fincham}. 
To account for the long range dipolar interaction $u_{DD}$, we apply the three-dimensional Ewald summation technique~\cite{klappbuch}. Specifically, we use a cubic simulation 
box with side length $L_x = L_y = L_z = L$ and employ periodic boundary conditions 
in a conducting surrounding. The parameter $\alpha$ which determines the convergence of the real space part of the Ewald sum is chosen to be
$\alpha = 6.0/L$ which is large enough to consider only the central box with $\boldsymbol{n} = 0$ in the real space. For the Fourier part of the Ewald sum we consider
wave vectors $\boldsymbol{k}$ up to $(\boldsymbol{k})^2=54$, 
giving a total number of wave vectors  $n_k = 1500$. In the MD simulations, we use the following
reduced units: $\rho^*= \, \sigma^3 \rho$, dipole moment $\mu^*=\sqrt{\epsilon\, \sigma^3}\mu$, time $t^*=\sqrt{\epsilon/(m\, \sigma^2)}t$ and temperature $T^*=k_BT/\epsilon$.
The simulations were carried out with $N=864$ particles and with a time step of $\Delta t^*=0.0025$. Typical simulations lasted for $3 \times 10^6$ steps.
%
%
%
\subsection{\label{simann}Simulated annealing}
To investigate ground state configurations of small clusters of particles interacting via the pair potential $u(i,j) = u_{HS} + u_{DD}$ [see Eqs.~(\ref{uDD}),~(\ref{uij}) 
and~(\ref{uhard})], we 
employ a simulated annealing procedure which involves a Monte Carlo simulation using the Metropolis algorithm \cite{tildesley}. Within this method, we choose initial 
states with comparable dipolar and thermal energies, i.e. $u_{DD}/k_BT \approx 1$. Here, $u_{DD}$ is the dipolar energy of two hard spheres in 
contact with central dipoles having head to tail orientation. We then lower the temperature stepwise to zero. At each temperature, 
$10^6$ trial moves are performed while conducting the usual Metropolis scheme involving translational
and rotational trial moves. We realize an acceptance
ratio of $60 \%$ by regularly adjusting the absolute value of the translational displacement during the simulation. New orientational configurations are generated by 
rotating the particles with a constant angle of $d\phi = \pi/18$ around one of the three axes of the laboratory fixed frame.
In order to ensure that we reach the state with lowest energy, we start several simulations for each set of parameters and choose those results 
with the lowest energy as the minimum energy state. \\
\section{\label{prelim}Ground state considerations} 
\subsection{Analytical expression for the pair energy}
\begin{figure} 
\begin{minipage}{9cm}
%
\includegraphics{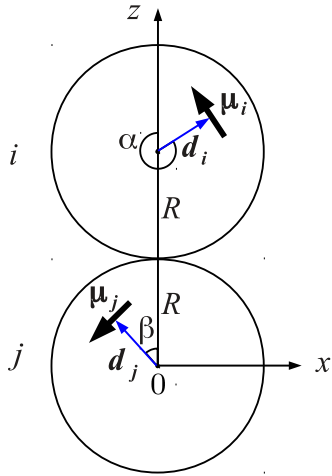} 
\end{minipage}
\caption{\label{skizze_analytisch}Sketch of two dipolar hard spheres $i$ and $j$ and the orientations of their shift and dipole vectors in the x-z-plane.}
\end{figure}

As a first step towards understanding the impact of the lateral shift, we consider two hard spheres with shifted dipoles (see Eq.~(\ref{uij}) with $u_{short}=u_{HS}$).
Specifically, we derive an analytical expression for the pair energies as function of the relative shift $\delta=|\boldsymbol{d}|/(2|\boldsymbol{R}|)$, where
$|\boldsymbol{R}|=\sigma/2$ is the particle radius. A similar derivation (leading to the same result) was very recently presented in~\cite{novak}. Here, we include the derivation 
as a background for later investigations of $N>2$ particles. The basis of 
the derivation is the coordinate system shown in Fig.~\ref{skizze_analytisch}. Note that this is a two-dimensional system ($x$-$z$-plane) where the orientations of the dipoles along the $y$-axis, i.e. out-of-plane orientations, are neglected. This assumption is 
confirmed by simulation studies of q2D dipolar systems showing that out-of-plane fluctuations vanish for decreasing temperatures~\cite{prokopieva}. In Fig.~\ref{skizze_analytisch},
the angles $\alpha$ and $\beta$ describe the orientations of the shift vectors $\boldsymbol{d}_i$ and $\boldsymbol{d}_j$ with respect to the $z$-axis. As the shift vector and 
the dipole vector have a fixed orthogonal orientation to each other, the orientations of $\boldsymbol{\mu}_i$ and
$\boldsymbol{\mu}_j$ with respect to the $z$-axis follow as $\alpha + \pi/2$ and $\beta + \pi/2$. With these definitions of the angles, the results for our lateral shift can be directly compared to those 
for the radial shift given in Ref.~\cite{holm2}.
Clearly, the distance $|\boldsymbol{r'}_{ij}|$ varies with $\alpha$, $\beta$ and $\delta$. Finally, we obtain
\begin{widetext}
\begin{equation} \label{udanalyt}
\begin{split}
u_{DD}(\delta,\alpha,\beta)&=\frac{\mu^2}{\sigma^3} \Biggl[\frac{\cos(\alpha-\beta) - 3\sin\alpha \sin\beta}{[\frac{\delta^2}{2}(1-\cos(\alpha-\beta))+1+\delta(\cos\alpha-\cos\beta)]^{3/2}}\\\\
&-\frac{3\delta^2\cos(\alpha+\beta)(\sin\beta-\sin\alpha)^2}{[\frac{\delta^2}{2}(1-\cos(\alpha-\beta))+1+\delta(\cos\alpha-\cos\beta)]^{5/2}}\\\\
&-\frac{3\delta(\sin\beta-\sin\alpha)(1+\delta(\cos\alpha-\cos\beta))\sin(\alpha+\beta)}{[\frac{\delta^2}{2}(1-\cos(\alpha-\beta))+1+\delta(\cos\alpha-\cos\beta)]^{5/2}} \Biggr] 
\end{split}
\end{equation}
\end{widetext}
for the dipolar potential $u_{DD}$ in terms of the parameters $\alpha$, $\beta$ and $\delta$. This expression is equivalent to that of Ref.~\cite{novak} 
(as can be seen after some rewriting.) 

\begin{figure}
\begin{center}
\scalebox{0.30}{\includegraphics{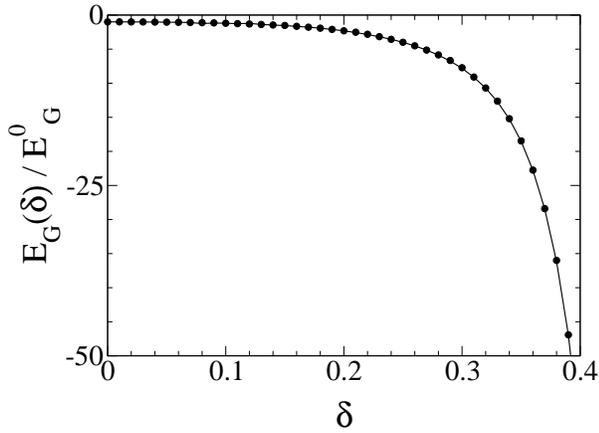}} 
\caption{\label{Eg_shift}Ground state pair energy $E_G$ normalized by the corresponding ground state energy $E_G^0 = -2{\mu}^2/\sigma^3$ of centred dipoles. The results 
are obtained by simulated annealing (circles) and 
by minimization of Eq.~(\ref{udanalyt}) (solid line).}
\end{center}
\end{figure}

%
We now aim at finding the minimum energy states, $E_{G}$, of two dipolar hard spheres as a function of $\delta$. To this end, we minimize Eq.~(\ref{udanalyt}) with 
respect to $\alpha$ and $\beta$ and compare the results with simulated annealing calculations in three dimensions, as described in Sec.~\ref{simann}. As the plot in 
Fig.~\ref{Eg_shift} clearly shows, the analytically gained results perfectly fit the numerical ones. Furthermore, it can be seen that the ground state energy $E_{G}(\delta)$
(which agrees with that calculated in~\cite{novak})
is a 
quantity which decreases with increasing shift. Initially, $E_{G}(\delta)$ changes slowly and is comparable to that of nonshifted dipoles suggesting that in this region, shifting the dipole moments out of the centres does not have a 
significant effect on the system. Upon further increase in $\delta$, $E_{G}(\delta)$ starts to rapidly decrease. This is a result of the fact that shifting the 
dipoles out of the centres enables them to reduce their distance compared to the case with zero shift. This effect becomes more and more pronounced with growing $\delta$ as 
the dipolar potential of Eq.~(\ref{uDD}) follows a power law of the dipolar distance.\\ 
When the results shown in Fig.~\ref{Eg_shift} are 
compared to the corresponding results of radially shifted dipoles of Ref.~\cite{holm1}, a qualitative agreement of the function $E_G(\delta)$ can be seen. Yet, 
in the case of lateral shifts, the reduction of energy sets in earlier,
i.e. for smaller shifts $\delta$ than those of radial shifts for which the energy starts to decrease only at $\delta \approx 0.25$. Further light on this issue is shed by 
inspecting the ground state configurations presented in the next section.
\subsection{\label{Gstructure}Ground state pair configurations}
\begin{figure}
\begin{center}                 
\scalebox{0.3}{\includegraphics{N2_angle_inset_grad.eps}}\\ 
\vspace*{0.5cm}
\quad\quad\scalebox{0.20}{\includegraphics{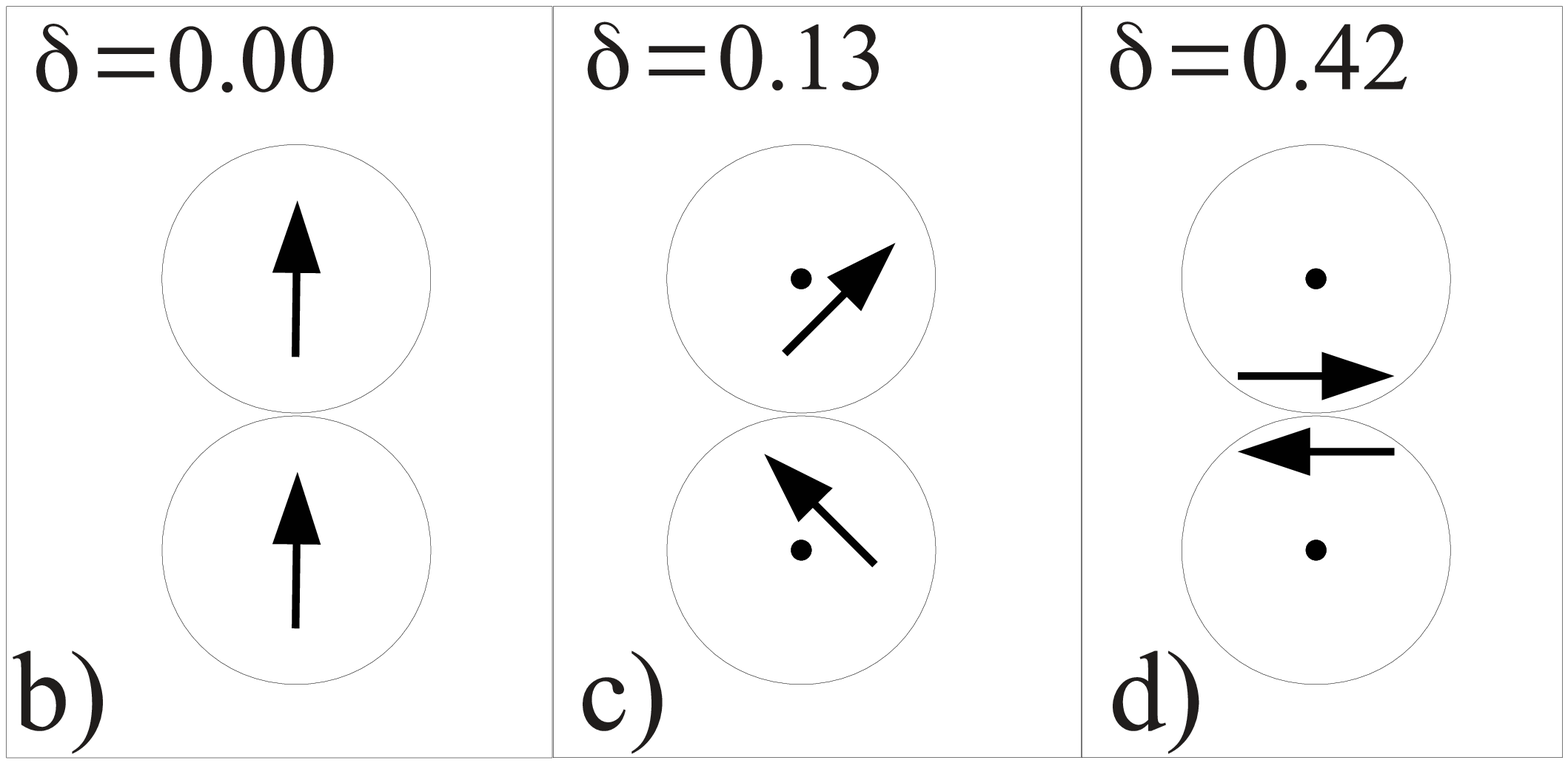}} 

\caption{\label{two_particles}(a) Simulated annealing results for the enclosed angle between the two dipoles in their ground states. The solid line indicates the zero line. 
The inset shows $\alpha$ and $\beta$ as defined in Fig.~\ref{skizzen} and gained by minimizing Eq.~(\ref{udanalyt}). (b)-(d) Ground state configurations of two dipoles.}
\end{center}
\end{figure}

The ground state configurations pertaining to a given shift $\delta$ are determined by those values for the angles $\alpha$ and $\beta$ that minimize Eq.~(\ref{udanalyt}). In 
Fig.~\ref{two_particles}, the angles $\alpha$ and $\beta$, as well as the cosine of the enclosed angle 
$\Delta(\delta) = \measuredangle (\boldsymbol{\mu}_1,\boldsymbol{\mu}_2)$ between the dipoles in their ground state arrangements at different shifts are shown. 
For $\delta = 0$, $\cos(\Delta = 0) = 1$ holds. We also note that for $\delta = 0$, the two sets $\alpha = \beta = \pi/2$ and $\alpha = \beta = 3\pi/2$ both
describe the ground state configuration (see Fig.~\ref{skizze_analytisch}) of nonshifted dipoles, which is the parallel head-to-tail orientation. Here, we choose the latter set of initial values, 
$\alpha = \beta = 3\pi/2$, as a starting point for our examination.\\  
Shifting the dipoles out of the centres, the parallel orientation of the dipoles is gradually abandoned in favour of reducing the dipolar distance. In detail, upon increasing
$\delta$ from zero, $\alpha$ 
is reduced until it reaches the value $\pi$ (see inset of Fig.~\ref{two_particles}a). Correspondingly, $\beta$ grows with increasing shift towards $2\pi$, as shown in the 
inset of Fig.~\ref{two_particles}. 
In other words, with increasing shift, the upper particle in Fig.~\ref{two_particles} (b) rotates clockwise while the lower one rotates counterclockwise and $\alpha$ and $\beta$ evolve in a completely 
symmetric manner. Thereby, $\Delta$ increases and $\cos(\Delta)$ decreases, reflecting that the dipoles more and more deviate from their parallel orientation. At the value 
$\delta \approx 0.13$, $\cos(\Delta)$ passes the zero line where $\Delta \approx \pi/2$ and the dipoles attain a perpendicular orientation. Finally, $\cos(\Delta)$ reaches its 
lowest value $\cos(\Delta) = -1$ (and thus $\Delta = \pi$) at 
$\delta = 0.2$. This corresponds to an antiparallel configuration of $\boldsymbol{\mu}_1$ and $\boldsymbol{\mu}_2$ relative to each other, 
and to a perpendicular orientation of each of the dipoles relative to the connecting line between the particle centres. For all higher 
shifts, the antiparallel orientation is kept and only the dipolar distance is further reduced. Interestingly, the value of $\delta = 0.2$ does not point any 
significance in the energy plot of Fig.~\ref{Eg_shift} but is highly significant for the preferred orientation of the dipoles. Thus we conclude that $\delta = 0.2$ represents 
the border between the two regimes of parallel (small shifts) and anti-parallel 
(high shifts) orientations (consistent with~\cite{novak}).\\
Compared to radially shifted dipoles of Ref.~\cite{holm1}, the main difference in the
ground state structures is that radially shifted dipoles keep their parallel head-to-tail orientation for small shifts. For large shifts, the two radially shifted dipoles
also attain an antiparallel oriented relative to each other whereas at the same time, each dipole is orientated along the connecting line between the centres of the 
particles. This demonstrates that not only the location but also the orientation of the dipole vector within the particle plays a crucial role for the ground states of the 
particles as also confirmed in Ref.~\cite{donaldson} in which the authors study the influence of shape and geometric anisotropy of the particles on their interaction.  
\subsection{\label{few}Triplet configurations and beyond}
\begin{figure} 
\begin{center}
\begin{minipage}[hbt]{9cm}
	\centering                                                         
	\quad\quad\scalebox{0.20}{\includegraphics{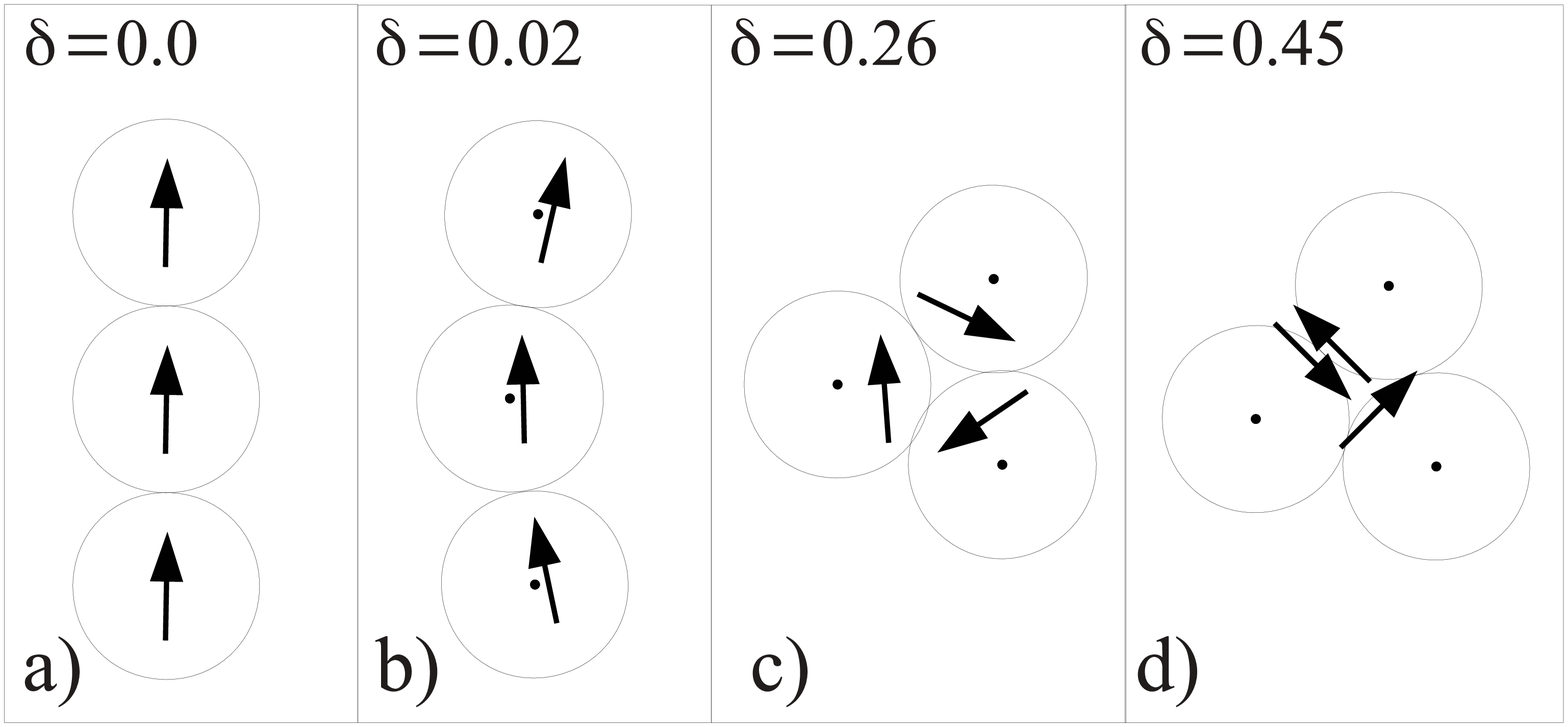}} \\
\end{minipage}
	\vspace*{0.5cm}  \\           
	\scalebox{0.25}{\includegraphics{N3angles_vs_shift_neu4.eps}} 
	\vspace*{-0.4cm}
\caption{\label{three_particles}(Color online) (a)-(d) Ground state configurations of three particles. (e) Simulation results for the three pair angles occuring between each of the three 
dipolar pairs in the ground state configurations. Each symbol represents one pair angle, respectively. For rectangular dipolar configurations, e.g at $\delta = 0.45$ [see (d)], 
two of the angles 
approach $90^{\circ}$ and the remaining one, accordingly, $180^{\circ}$.}
\end{center}
\end{figure}

\begin{figure}
	
	\quad\scalebox{0.3}{\includegraphics{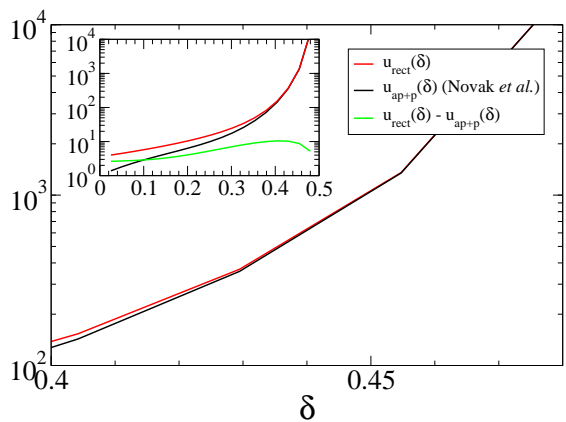}} 
\caption{\label{vgl_energies}Absolute values for $u_{rect}(\delta)$, $u_{ap+p}(\delta)$~\cite{novak} and for the difference $|u_{rect}(\delta)| - |u_{ap+p}(\delta)|$, in a.u., respectively.}
\end{figure}

\begin{figure}
	\vspace*{0.7cm} 
	\quad\scalebox{0.2}{\includegraphics{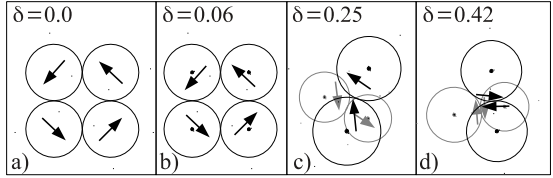}} 
	\vspace*{-1cm}
\caption{\label{four_particles} Ground state configuration of four particles for several shifts $\delta$.}
\end{figure}

The principle impact of shifting the dipoles out of the particles' centres, namely, the decrease of the ground state energy and a preferred non-parallel orientation
of the dipoles with increasing shift, becomes even more pronounced in systems of three and four dipolar hard spheres.\\
For a detailed investigation, we have performed simulated annealing calculations to determine the ground state configurations of three-dimensional systems with three and four 
hard spheres for different shifts. We first consider the three-particle case.
In Fig.~\ref{three_particles}(a)-(d), we sketch the obtained configurations. 
Starting from the chainlike head-to-tail orientation known for nonshifted 
dipoles, the particles first organize into slightly curved chainlike geometries [Fig.~\ref{three_particles}(b)]. This occurs for very small 
shifts up to $\delta \approx 0.03$. Our simulation results show that the corresponding ground state energies for this curved chain configuration is indeed 
slightly lower (see Table~\ref{energy_table}) than those 
for the corresponding structure proposed in Ref.~\cite{novak} which the authors call a "zipper". In a "zipper" configuration the dipoles 
have head-to-tail orientation and are organized in a staggered manner.\\ 
When the shift takes values above $\delta \approx 0.03$, the two particles at 
the ends of the chain approach each other in such a way that they form a planar triangular arrangement. This behavior remains for all higher shifts 
[Fig.~\ref{three_particles}(c) and (d)], in agreement with the results of 
Ref.~\cite{novak}.\\ In terms of the orientations of the dipoles within the 
particles, in the case of chainlike geometries, the dipoles show
head-to-tail orientation. Within the triangular geometries, there are two qualitatively different types of dipolar orientations. The first type is likewise triangular with 
all the pair angles (i.e. the angles between the three dipolar pairs) attaining the value of $120^{\circ}$ for shifts up to $\delta \approx 0.38$, as confirmed by the plot in Fig.~\ref{three_particles}(e). The second type is 
a rectangular orientation in which two of the dipoles form an antiparallel pair and the third one joins the pair in a perpendicular manner [Fig.~\ref{three_particles}(d)]. 
Correspondingly, two of the three pair angles have a value of $90^{\circ}$ and the third one of $180^{\circ}$, as shown in Fig.~\ref{three_particles}(e). 
We note that at higher shifts $\delta \agt 0.4$, we find again a difference to the results in~\cite{novak}. The authors propose a configuration containing 
an antiparallel pair which is joined by the third particle via a head to tail orientation with one of the dipoles of the antiparallel pair. To clarify this issue, we have derived 
an analytical expression for the rectangular configuration in Fig.~\ref{three_particles}(d). It is given by
\begin{center}
\begin{eqnarray*} \label{urect}
u_{rect}(\delta) = -\frac{1}{(1-2\delta)^3} -\frac{3(1-2\delta)(\frac{\sqrt{3}}{2}-\delta)}{\sqrt{(1+2\delta^2-\delta(1+\sqrt{3}))}^{\,5}}\, .
\end{eqnarray*}
\end{center}
Evaluating this energy, we find that the rectangular configuration is energetically slightly more favourable than that of Ref.~\cite{novak}. 
Figure~\ref{vgl_energies} shows the results for the absolute values of $u_{rect}(\delta)$, the results for the absolute values of Eq.~(7) of Ref.~\cite{novak}, 
$u_{ap+p}(\delta)$, and the difference $|u_{rect}(\delta)| - |u_{ap+p}(\delta)|$, which is positive for all values considered. \\
Finally, in the case of four particles, the nonshifted ground state configuration is a ring geometry with rectangular, cyclic orientation of 
the dipoles, as it is known from other ground state studies~\cite{holm2} [see Fig.~\ref{four_particles}(a)]. This configuration 
remains for small shifts where only the dipolar distances are reduced while the orientations are maintained. Upon increasing the shift, opposing dipoles within the rectangular 
geometry more and more approach each other and form two pairs of antiparallel dipoles which are perpendicular oriented to each other. This is accompanied by a change from
the planar rectangular towards a tetrahedral configuration as shown in Fig.~\ref{four_particles}(c) and (d). Thus, in the four particle system, we observe for the first time 
a cross-over from planar to three-dimensional configurations.\\
%
%
%
%
%
%
\begin{table}
\caption{\label{energy_table}Ground state energies $E_{gs}$ in a.u. for three dipolar hard spheres gained by simulations for very small shifts. The corresponding ground state configurations are sketched 
 in Fig.~\ref{three_particles}(b) }
 \begin{ruledtabular}
  \begin{tabular}{ll}
  $\delta$ & $E_{gs}$ in a.u.\\
   
    0.0125 & -4.2556 \\ 
    0.01875 & -4.2667 \\ 
    0.02 & -4.2722 \\ 
    0.025 & -4.2911 \\
 \end{tabular}
 \end{ruledtabular}
\end{table}

 
%
%
\section{\label{bulk}Systems at finite temperature}
\subsection{Preliminary considerations}
In this section, we investigate finite temperature systems with soft-sphere repulsive interactions, which seem more realistic for 
the real colloidal particles mentioned in the introduction. To this end, we set in Eq.~(\ref{uij}) the parameters $\epsilon = 50$ and $n=38$.\\
Due to the fact that the magnitude of the ground state energy $E_{G}(\delta)$ is an increasing function of the shift (see previous discussions), also the dipolar 
coupling strength $\lambda$, which is defined as the ratio of the half ground state energy and the thermal energy, $\lambda(\delta) = |E_{G}(\delta)|/2k_BT$,
becomes an increasing function of the shift. This yields an irreversible agglomeration of the particles, which cannot be counteracted by the soft-core potential.
For the present choices for $\epsilon$ and $n$ this situation occurs if the shift exceeds the value of $\delta = 0.33$. We examined higher shifts than $\delta=0.33$ by 
appropiate choices for $\epsilon$ and $n$ but did not gain any new insights of the system beyond those already observed for smaller shifts. Therefore, instead of adjusting $\lambda(\delta)$, e.g by appropriate reduction of $\mu^*$ 
with increasing shifts, or instead of enhancing the soft-sphere potential values $\epsilon$ and $n$, we limit the shift at 
$\delta_{limit}=0.33$ in order to prevent agglomeration. In this way the structural properties of the system can be directly related to the amount of shift which hence is the 
parameter of interest in our examinations.\\
We consider a strongly coupled system with $\mu^* = 3$ with the densities $\rho^*=0.07$, $\rho^*=0.1$ and $\rho^*=0.2$ and at the two temperatures $T^*=1.0$ and $T^*=1.35$, 
respectively. This yields coupling strengths ranging from $\lambda(\delta = 0) = \mu^2/(k_BT\, \sigma^3) = 9$ to $\lambda(\delta = 0.33) = 72$ for $T^*=1.0$, and 
$\lambda(\delta = 0) \approx 6.67$ to $\lambda(\delta = 0.33)\approx 53.33$ for $T^*=1.35$. 
For a thorough investigation of the equilibrium properties of the shifted system, we performed MD simulations and calculated various structural properties, as described
in the next section. 
\subsection{\label{rdf} Results} 
\begin{figure}                         
	\centering\scalebox{0.45}{\includegraphics{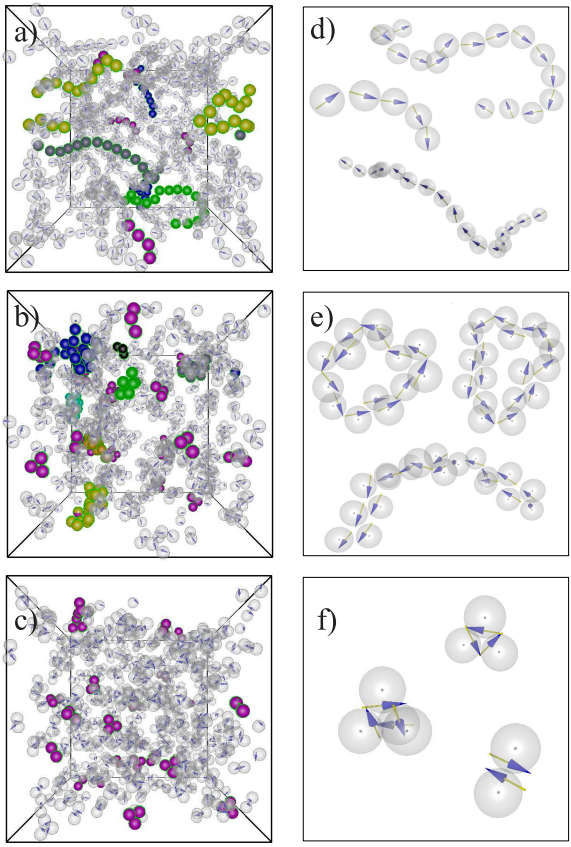}} 
\caption{\label{snapshots} (Color online) Snapshots for $\delta=0$ (a), $\delta=0.21$ (b) and $\delta=0.33$ (c) with revealing structures of the group (A) (a), 
the groups (B), (C) and (D) (b) and only (D) (c). In each snapshot, some randomly chosen
clusters are coloured for a better visibility. Particles of the same color besides magenta belong to the same cluster. Magenta colored clusters represent small single 
clusters (D). (d)-(f) 
Magnification of randomly chosen clusters of the snap shots in the left column.} 
\end{figure}

%
For a first overview, we present in Fig.~\ref{snapshots} representative MD simulation snapshots illustrating typical self assembling structres. Specifically, we consider 
systems at $T^*=1.0$ and $\rho^*=0.1$ for $\delta = 0$, 
$\delta = 0.21$ and $\delta = 0.33$ . \\ 
Qualitatively, the structures appearing for the considered values of $\delta$ can be divided into four groups. These are chains (A), staggered chains (B), rings built by 
staggered chains (C) and small clusters (D) of the types presented in Figs.~\ref{two_particles}(d),~\ref{three_particles}(c) and~\ref{four_particles}(c). 
Structures of
type (A) can consist of a few ($2-5$) as well as of many (more than $10$) particles, i.e., the chains can be short or long. Structures of types (B) and (C) always consist of
more than $10$ particles [Fig.~\ref{snapshots}(d), (e)].
In accordance with the ground state configurations (see Figs.~\ref{three_particles} and~\ref{four_particles}), the structures found in the finite temperature systems for different shifts pass from 
chainlike geometries to circular close-packed clusters upon the increase of $\delta$. 
Accordingly, structures of the first group are formed for zero and small shifts in the range $\delta=0.01 - \delta \approx 0.1$ [Fig.~\ref{snapshots}(a) and (d)]. In this 
shift region, the overall chainlike structure with head-to-tail orientation as formed by nonshifted dipoles is maintained. Yet, the shift causes more and more curved 
structures compared to the nonshifted particles. As is generally known for dipolar systems, the chain length, i.e. the number of particles within a chain, has a polydisperse
distribution~\cite{teixeira}. This holds also for the shifted system (see also the discussion of the cluster analysis in Sec.~\ref{cluster}).\\ 
For intermediate shifts, e.g. $\delta = 0.24$, Fig.~\ref{snapshots}(b) and (e), the particles within the chains become staggered and we observe 
coexistence of structures of the types (B), (C) and (D). Structures of group (D) are consistent with ground the state configurations of this and higher shifts. Although 
groups (B) and (C) are not observed for zero temperature, they can be understood as a modification of chains, as they appear for small $\delta$, and of rings which occur 
at zero temperature.\\ 
If $\delta$ takes values near $0.33$, all large aggregates (B) and 
(C) vanish and only small clusters (D) remain, as shown in Figs.~\ref{snapshots}(c) and (f).\\
The same structural behaviour at the different shift regions is observerd for the other state points considered. Thus we conclude that the described self-assembly of the 
particles at different shifts is a quite general behaviour which results from the increasing dipolar coupling strength for increasing shifts. The latter causes 
more and more close-packed structures as we already confirmed in the case of hard spheres.\\
%
%
%
%
%
%
\subsubsection{Radial distribution function}
\begin{center}       
\begin{figure}                
	\includegraphics[width=8.3cm]{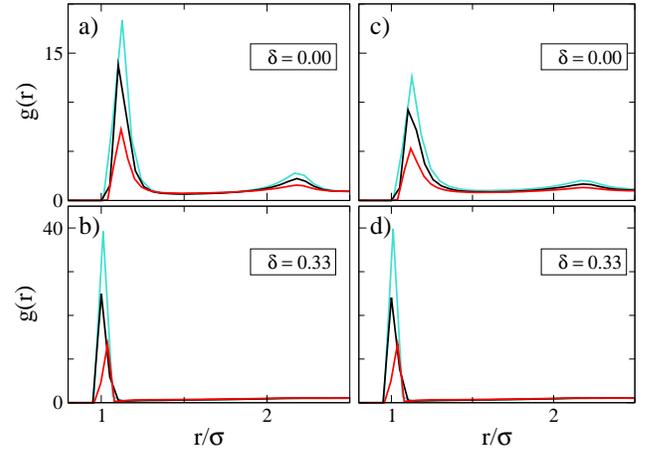}
\caption{\label{figsrdf} (Color online) Radial distribution functions $g(r)$ for densities $\rho^* = 0.07$ (turquoise), $\rho^* = 0.1$ (black) and $\rho^* = 0.2$ (red) at 
two temperatures $T^* = 1.0$ in (a) and (b), and $T^* = 1.35$ in (c) and (d).}
\end{figure}
\end{center}

As a first quantitative measure of the structure formation, we consider the radial distribution function \\
$$g(r) = \frac{\langle \sum_{i \ne j}\delta(r-r_{ij}) \rangle}{N\rho 4 \pi r^2}$$\\ for several shifts.\\
The plots in Fig.~\ref{figsrdf} show $g(r)$ for $\delta = 0$ and $\delta = 0.33$ for $T^*=1.0$ and $T^*=1.35$. The $g(r)$ at zero shift is dominated by first and second
neighbour correlations. This is a typical feature of strongly coupled dipolar systems~\cite{weipre,klapppaper} and reflects the formation of 
chain-like structures. When we successively increase the shift, the second peak exists up to a value of $\delta \approx 0.25$. Beyond this value, only nearest neighbour 
correlations at $r/\sigma=1$ are present in the system signifying the presence of 
only small and close-packed clusters (D), as seen in the snap shots of Fig.~\ref{snapshots}(c).\\
Noticeably, the results for the higher temperature $T^*=1.35$ completely coincide with 
those of $T^*=1.0$ in the high shift region [Fig.~\ref{figsrdf}(b) and (d)]. This is because for sufficiently high shifts, 
the increase of the dipolar coupling strength is already enhanced and thus, the increase of temperature does not affect the self-assembly.
%
%
%
%
\subsubsection{\label{cluster}Cluster analysis}
To further characterize the aggregates, we perform a cluster analysis. In particular, we are interested in the cluster size distribution for several shifts, the mean cluster 
size and the mean cluster magnetization as a function of $\delta$. The basis of this analysis are
distance and energy criteria. Specifically, all particles with a distance lower than $r_c=1.3\,\sigma$ and binding energy 
$u_c = \sum_{i,i'>i} u^{ii'}_{DD} < 0$ are regarded as being clustered. Here, $u^{ii'}_{DD}$ denotes the dipolar energy [see Eq.~(\ref{uDD})] between all pairs
$i,i'$ within the critical distance $r_c$.\\
The detected clusters were collected in a histogram in which the number of clusters with size $S$, $n(S)$, is counted and normalized by the total number of clusters,
$N_c = \sum_{S \ge 2} n(S)$, such that
$$h(S) = \Big \langle \frac{n(S)}{N_c} \Big \rangle\, ,$$ gives the normalised cluster size distribution. 
Only $S \ge 2$ enters to the sum, i.e., single particles are disregarded.\\
\indent Based on the function $n(S)$, the mean cluster magnetization is calculated by\\
$$\langle M \rangle = \Big \langle \frac{\sum_{S \ge 2} n(S) \cdot M_c(S)}{N_c} \Big \rangle \, ,$$\\ where 
$M_c(S) = \Big |\sum_{i=1}^{S} \boldsymbol{\mu}_i/(\mu\cdot S) \Big |$ gives the normalized magnetization of a cluster with size $S$. The quantity $M_c(S)$ 
is a measure of parallel allignment of the dipole vectors within the individual clusters. Specifically, values of $M_c(S)$ near to one reflect a high degree of head to tail 
orientation,
while vanishing values of this quantity indicate antiparallel or triangular orientation. Therefore, the mean cluser magnetization gives inside
into the organization of the dipoles within the formed structures and thus allows to evaluate if a given assembly is chainlike [types (A) and (B)] or closed 
[types (C) and (D)]. Note that the total magnetization, which is usually calculated by summing 
over all particles, has vanishing values as the system is globally isotropic at the state points considered here.\\  
Finally, 
the mean cluster size is obtained from\\
$$\langle S \rangle = \Big \langle \frac{\sum_{S \ge 2} n(S) \cdot S}{N_c} \Big \rangle\, . $$\\ 

\begin{center}       
\begin{figure}                
	\includegraphics[width=8.3cm,height=6.5cm]{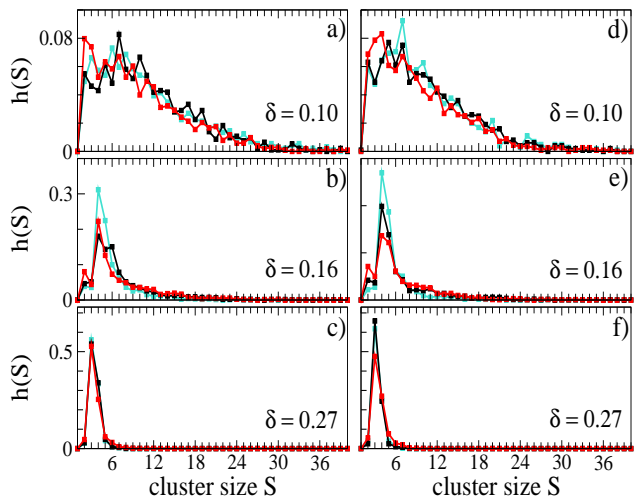}
\caption{\label{figsdists} (Color online) Normalized cluster size distribution for the same densities and colors as in Fig.~\ref{figsrdf}. (a)-(c): $T^* = 1.0$. (d)-(f): $T^* = 1.35$.}
\end{figure}
\end{center}

{\it{(a) Normalised cluster size distribution}}. The results for $h(S)$ for different characteristic shifts, 
namely for $\delta = 0.1$ (small shift), $\delta = 0.16$ (intermediate shift) and
$\delta = 0.27$ (high shift) are presented in Fig.~\ref{figsdists}. 
The figures~\ref{figsdists}(a) and (d) show that mostly large aggregates, that can contain up to $25 - 30$ particles, are 
formed. On the other hand, Figs.~\ref{figsdists}(c) and (f) indicate the formation of only small assemblies with $3-4$ particles.\\
However, in Fig.~\ref{figsdists}(b) and (e),
although there is a preferential emergence of small assemblies, large aggregates of up to $20$ particles are present in a non-negligible number and 
secondary peaks at e.g. $S = 15$ (for $T^*=1.0$) and $S=13$ (for $T^*=1.35$) are visible. Evidently, for this and comparable shifts, small and large assemblies can coexist.\\
One also finds that for higher temperature, large aggregates are less often formed than for the smaller 
temperature. This is indicated by the fact that the peaks in Figs.~\ref{figsdists}(e) and (f) are enhanced compared to those in 
Figs.~\ref{figsdists}(b) and (c).\\

\begin{center}       
\begin{figure}   
    \vspace*{0.5cm}
    \includegraphics[width=8.3cm,height=6cm]{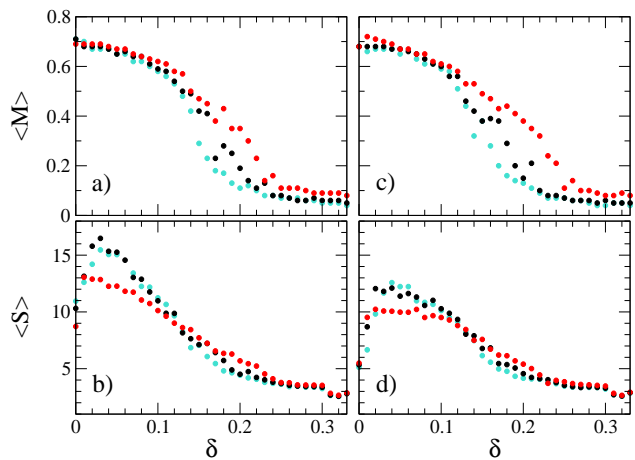}
\caption{\label{figs_size_mag} (Color online) Mean cluster size $\langle S \rangle$ and mean cluster magnetization $\langle M \rangle$ as a function of the shift at two temperatures
$T^* = 1.0$ ((a),(b)) and $T^* = 1.35$ ((c),(d)). Colors are the same as in Fig.~\ref{figsrdf}. }
\end{figure}
\end{center}

{\it{(c) Mean cluster magnetization}}. In order to evaluate the types of the occuring structures for a given shift, we determine $\langle M \rangle$ as a
function of the shift and plot the results in Figs.~\ref{figs_size_mag}(b) and (d).\\
For zero and initial shifts, $\langle M \rangle$ takes the value $\approx 0.7$, 
reflecting predominantly parallel orientation of the dipoles within their aggregates. From this and from the cluster size distribution  
[Fig.~\ref{figsdists}(a),(d)] we conclude that for small shifts (up to $\delta \approx 0.1$), mainly short and long polar chains of type (A) or (B) are formed.\\ 
If the shift is further increased, $\langle M\rangle$ decreases, indicating that polar chains occur less often. Instead, the aggregates become more and more closed 
structures of the types (C) or (D) with increasing shifts. Hence, the decrease of 
$\langle M \rangle$ implies the coexistence of types (B), (C) and (D) [see Figs.~\ref{snapshots}(b) and (e)]. At the high shift end, $\langle M\rangle$ drops down to 
vanishing values indicating only pairwise antiparallel or triangular arrangements of the dipoles within the clusters, which is also consisent with the results shown in 
Fig.~\ref{figsdists}(c) and (f). The fact that the mean cluster magnetization has vanishing values at large $\delta$ also suggests that the clusters poorly 
interact.\\
Note that for all values of $\delta$, the according aggregates are isotropically 
oriented such that the total magnetization is zero for all shifts (not shown here).\\

{\it{(b) Mean cluster size}}. Finally, we examine the influence of the shift on the mean cluster size and plot in Figs.~\ref{figs_size_mag}(a) and (c) 
$\langle S \rangle$ as a function of the shift.\\
Starting at $\delta = 0$, the mean cluster size grows to its maximum with about $17$ particles for $T^* = 1.0$ and about $13$ particles for $T^* = 1.35$. The maximum is 
reached at $\delta \approx 0.05$, respectively. This increase can be understood by the effective increase of the dipolar coupling strength $\lambda$ (see preceding 
discussion) such that initial shifts result in the formation of longer chains of type (A). If $\delta$ exceeds this value, $\langle S\rangle$ starts to gradually decrease 
because with increasing shift, smaller aggregates are formed more frequently (see Fig.~\ref{figsdists}). Finally, $\langle S\rangle$ attains the value of about $3$ particles 
in the high shift end, which is a highly representative value for both temperatures considered [Figs.~\ref{figsdists}(c) and (f)]. Significant differences between the results
of the two temperatures can be seen only
for shifts smaller than $\delta \approx 0.1$ where mainly chainlike aggregates are formed. Here, the increase of temperature, 
which involves the decrease of the coupling strength from $\lambda = 9$ to $\lambda \approx 6.67$, causes the formation of 
chains with less particles. 
Moreover, for these values of $\delta$, shifting the dipoles does not impose fundamentally different self-assembly 
patterns compared to nonshifted dipoles. Therefore, small shifts can be regarded as perturbation of the nonshifted system.\\
On the other hand, high shifts impose significantly different structures: the particles exclusively form structures of type (D) that 
correspond to ground state configurations of two, three and four hard spheres [see Figs.~\ref{two_particles}(d),~\ref{three_particles}(c) and~\ref{four_particles}(c)]. 
This is possible due to the large values of $\lambda = 72$ for $T^*=1$ or $\lambda \approx 53.33$ for $T^*=1.35$.\\
Finally, for intermediate shifts, where large aggregates as well as small clusters are formed, the decrease of $\langle S\rangle$ (and at the same time of $\langle M \rangle$)
can be interpreted as a transition region in 
which large aggregates gradually dissolve into small clusters until no large structures appear at all. Within this region, the competition between energy minimization and 
entropy maximization results in the coexistence of both,
small and large aggregates. With increasing shift (i.e., effectively increasing $\lambda(\delta)$), the particles accomplish to form structures equivalent 
to ground state configurations.\\

To summarize, in the bulk systems at the finite temperatures and densities considered here, we can qualitatively distinguish between three shift regions 
(small, intermediate and high) each of which is characerized by it's own structural characteristics.
By contrast, in the ground states of two particles, we determined only a small and a high shift region (see Fig.~\ref{two_particles}(a) 
and the related discussion). The intermediate shift region, observed for the bulk systems is not detected for zero temperature. This is consistent with the fact that the 
corresponding structures of types (B) and (C) are not observed in the ground state calculations. \\
\section{\label{conclusion}Conclusions and outlook}
In this paper, we investigated a model of spherical particles with laterally shifted dipole moments which is inspired by real micrometer sized particles that carry a magnetic component on 
or right beneath their surfaces (e.g. \cite{granick,bibette,pine}). 
Aiming at understanding
the principle impact of the shift of the dipole moment on the three-dimensional system, we determined the ground state structures 
of two, three and four dipolar hard spheres. It turns out that shifting the dipole fundamentally affects ground state energies and 
configurations, as 
well as self-assembly patterns in finite
temperature systems. For these, we could determine three regions 
of shift, being small, intermediate and high. In each region, the self-assembly of the particles is fundamentally different. The system passes from a state which is similar 
to that of nonshifted dipoles to a clustered structural state. 

Further, it is an interesting observation that the asymmetry of the particles, caused by the off-centred location of the dipole moment, is overcome for small shifts 
insofar as the behaviour of the small shift region can be recognized as a perturbation of the nonshifted system.
On the other hand, if the shift is too high, the system compensates the off-centred location of the dipole by building symmetric aggregates.

So far, we examined the equilibrium properties of systems of shifted dipoles. Further investigations should clarify the interaction between the aggregates
in the different shift regions. Moreover, it would be desirable to have a full phase diagram as it is known for centred dipolar soft spheres 
\cite{wei}.\\

In view of the severe effects of the shift on the equilibrium properties, one expects new types of pattern formation if the system is out of equilibrium.
An interesting case are systems of shifted dipoles exposed to several types of external magnetic fields. The case of a constant field was examined
in Ref.\cite{novak} demonstrating that shifted dipoles form staggered chains for appropriate values for the field strength and the shift. Even more exciting phenomena are 
provoked if the field is time-dependent, e.g. precessing or rotating, driving the particles accordingly into tubular~\cite{granick} or crystalline structures~\cite{yan} 
as a result of synchronization effects in the systems.
The immediate interest particularly lays in the question to which extent the model of laterally shifted permanent dipoles can be used to theoretically 
describe phenomena observed in real systems such as Janus particles~\cite{granick,yan}. Computer simulations in these directions are on the way.
\section*{\label{acknowledgement}Acknowledgements}
We gratefully acknowledge financial support from the DFG within the research training group RTG 1558 {\it{Nonequilibrium Collective Dynamics in Condensed Matter and 
Biological Systems}}, project B1. We also thank Rudolf Weeber and Christian Holm for discussions related to the derivation of the equations of motion of the laterally
shifted dipoles.

\end{document}